\documentclass[aps,prl,twocolumn,groupedaddress,showpacs]{revtex4}
\usepackage{epsfig}

\def\k{{\bf k}}

\def\8{\infty}

\def\undertext#1{\vtop{\hbox{#1}\kern 1pt \hrule}}

\def\pp#1{\frac{\partial}{\partial#1}}
\def\pbyp#1#2{\frac{\partial#1}{\partial#2}}

\def\be{\begin{equation}}
\def\ee{\end{equation}}
\def\bea{\begin{eqnarray} & &}
\def\eea{\end{eqnarray}}

\def\rf#1{(\ref{#1})}

\def\rf#1{(\ref{#1})}

\def\r{{\bf r}}

\def\rfs#1{Eq.~\rf{#1}}

\begin{document}


\title{Zero modes of two-dimensional chiral $p$-wave  superconductors}


\author{V. Gurarie}
\author{L. Radzihovsky}
\affiliation{Department of Physics, University of Colorado,
Boulder CO 80309}


\date{\today}

\begin{abstract}
  We discuss fermionic zero modes in the two-dimensional chiral $p$-wave
  superconductors. We show quite generally, that without fine-tuning,
  in a macroscopic sample there is only one or zero of such
  Majorana-fermion modes depending only on whether the total vorticity
  of the order parameter is odd or even, respectively. As a special
  case of this, we find explicitly the one zero mode localized on a
  single odd-vorticity vortex, and show that, in contrast, zero modes
  are absent for an even-vorticity vortex. One zero mode per odd
  vortex persists, within an exponential accuracy, for a collection of
  well-separated vortices, shifting to finite $\pm E$ energies as two
  odd vortices approach. These results should be useful for the
  demonstration of the non-Abelian statistics that such zero-mode
  vortices are expected to exhibit, and for their possible application
  in quantum computation.
\end{abstract}

\maketitle

Recently \cite{Read2000,Nayak2006a,Tewari2006} there has been
considerable interest in the structure of fermionic zero modes
localized on vortices of a chiral spinless two-dimensional
superconductor characterized by $p_x+ip_y$ order parameter. In part, it is stimulated by a proposal
\cite{Read2000} that a ground state of such a superconductor (for a
positive chemical potential) is
similar to the Moore-Read (Pfaffian) quantum Hall state \cite{MR},
thought to describe the $\nu=5/2$ quantum Hall plateau
\cite{Willett1987}. Vortices (corresponding to the Laughlin
quasihole-like excitations \cite{Read1996} in the Moore-Read state) in
such a superconductor are thus expected to exhibit a degenerate set of
zero modes separated from all other states by a gap, and to obey
non-Abelian statistics \cite{Ivanov2001}, that may make them useful
for a realization of a ``topological quantum computer''
\cite{Kitaev2003} free of decoherence.

Many of the properties of these zero modes for a {\em single} vortex
have already been discussed in the
literature~\cite{Read2000,Nayak2006a,Tewari2006}.  However, in our
view an explicit discussion of the fate and robustness of the zero
modes to, for example, a local deformation of the order parameter or
in the presence of many vortices has not appeared in the literature.
Such questions are of particular interest in view of recent proposals
for experimental realization and manipulation of such non-Abelian
states in two-dimensional superconductors, such as Sr$_2$RuO$_4$
\cite{Nayak2005a}, the $\nu=\frac 5 2$ plateau of the quantum Hall
effect \cite{Stern2006,Shtengel2006,Feldman2006}, and $p$-wave
resonantly interacting atomic superfluids
\cite{Gurarie2005,Nayak2006a}.

In this Letter we show quite generally, that for a macroscopic sample
(i.e., ignoring the boundary physics), without fine-tuning, strictly
speaking there is only {\em one} or {\em zero} Majorana-fermion mode
depending only on whether the total vorticity of the order parameter
(in elementary vortex units of $2\pi$) is {\em odd} or {\em even},
respectively.  For a collection of well-separated vortices, within an
exponential accuracy one zero mode per an isolated odd-vorticity
vortex persists. As two of such vortices are brought closer together
the corresponding pair of ``zero'' modes splits away to finite $\pm E$
(vortex-separation dependent) energies.  Generically, even-vorticity
vortices do not carry any zero modes.

Even in the odd-vorticity case, zero modes only exist for a {\em
  positive} chemical potential $\mu>0$, consistent with the existence
(absence) of a topological order in a weakly- (strongly-) paired
ground state of a $p$-wave superconductor stable only for $\mu>0$
($\mu<0$)~\cite{Read2000,VolovikBook}.  While a $p$-wave superconductor in a solid
state context naturally obeys $\mu>0$, in a Feshbach resonant atomic
$p$-wave superfluid a chemical potential can be adjusted to be
positive via an external magnetic field~\cite{Gurarie2005}, a ``knob''
that can also be used to drive a topological quantum phase transition
between a strongly- and weakly-paired superfluid ground states~\cite{Read2000}.

As a demonstration of a specific realization of this general
connection between parity of vorticity and a number of zero modes, we
compute the eigenfunction of the one zero mode localized on a single
isolated odd-vorticity vortex, and show that zero modes are absent for
an even-vorticity vortex. This symmetric vortex result is in agreement
with a recent study in Ref.~\cite{Tewari2006}, but does not rely on a
linearization of the fermion dispersion around a Fermi surface, and
thereby allows us to access the nondegenerate (low chemical potential)
regime realizable in tunable (via a Feshbach resonance
\cite{Gurarie2005}) atomic gas experiments.  Our results then imply
that such zero-modes, residing on isolated elementary vortices are
always shifted to finite $\pm E$ energies when an even number of them
come into proximity \cite{Read2000}, with possible deleterious
implications for a realization of non-Abelian statistics and quantum
computation.

To demonstrate these results we begin by first discussing the
properties of the solutions of generic Bogoliubov-de-Gennes (BdG)
equations arising in a context of any superconductor. These coupled
Schr\"odinger equations follow from the following
Bardeen-Cooper-Schrieffer (BCS) Hamiltonian
\begin{equation}
H = \sum_{ij} \left( a^\dagger_i h_{ij} a_j -a_j h_{ij} a^\dagger_i+
a_i \Delta_{ij} a_j + a^\dagger_j \Delta^*_{ij} a^\dagger_i
\right),
\label{Ha}
\end{equation}
where indices $i$, $j$ label space (and in a spinful case, spin)
coordinates of the fermion creation and annihilation operators
$a^\dagger_i$, $a_i$.  Their canonical anticommutation relations
ensure that $\Delta_{ij}$ is an antisymmetric operator. Since $H$ must
be hermitian, so is $h_{ij}$, and the problem is equivalent to a study
of the spectrum and eigenstates of a matrix
\begin{equation} \label{eq:classD}
{\cal H}  = \left( \matrix { h & \Delta \cr \Delta^\dagger & - h^T}
\right).
\end{equation}
This matrix possesses the following important symmetry property
\begin{equation} \label{eq:symmetryD}
\sigma_1 {\cal H} \sigma_1 = - {\cal H}^*,
\end{equation}
where $\sigma_1$ is the first Pauli matrix acting in the 2 by 2 space
of the matrix ${\cal H}$, \rfs{eq:classD}. In the terminology of
Ref.~\cite{Altland1997}, the matrix ${\cal H}$ is said to belong
to the symmetry class $D$.  As a result of this property, it can be
seen from
\begin{equation}
{\cal H} \sigma_1 \psi^* = - \sigma_1 {\cal H}^* \psi^*=-E \sigma_1
\psi^*
\end{equation}
that if $\psi$ is an eigenvector of such ${\cal H}$ with the
eigenvalue $E$, then $\sigma_1 \psi^*$ is guaranteed to be an
eigenvector with the eigenvalue $-E$.  As a result, all nonzero
eigenvalues of ${\cal H}$ come in $\pm E$ pairs. A special role is
played by the zero eigenvectors of this matrix, referred to as 
zero modes. If $\psi$ is a zero mode, $\sigma_1 \psi^*$ is also a zero
mode. Taking 
linear combinations $\psi+\sigma_1 \psi^*$, $i(\psi-\sigma_1 \psi^*)$
of these degenerate modes, we can always ensure the
relation
\begin{equation}
\label{eq:rela}
\sigma_1 \psi^*=\psi
\end{equation}
for every zero mode. In the absence of other symmetries of ${\cal H}$
it is quite clear that generically there is nothing that protects the
total number $N_z$ of its zero modes under smooth changes of the
Hamiltonian matrix that preserve its BdG form. However, since non-zero
modes have to always appear in $\pm E$ pairs, as long as the symmetry
property \rf{eq:symmetryD} is preserved by the perturbation the number of
zero modes can only change by multiples of $2$.  Thus, while the
number $N_z$ of zero modes of the Hamiltonian \rf{eq:classD} may
change, this number will always remain either odd or even, with
$(-1)^{N_z}$ a ``topological invariant''
\cite{ZirnbauerPrivate,NickReadUnpublished}.

The value of this invariant is easy to establish if one observes that
${\cal H}$ is an even sized matrix, with an even number of
eigenvalues. Since the number of non-zero modes must be even, this
implies, quite generally that the number of zero modes is also even,
$(-1)^{N_z}=0$, and strictly-speaking the BdG Hamiltonian does not have any
topologically protected zero modes.  Furthermore, since zero modes must appear in pairs, there can only be
an even number of accidental zero modes, which will nevertheless be
generally destroyed by any perturbation of ${\cal H}$ (preserving its
BdG structure \rfs{eq:classD}).  We believe this observation was first
made by N. Read~\cite{NickReadUnpublished}.

The situation should be contrasted with that of the Dirac operators
${\cal D}$. Those operators, being generally of one of the chiral
classes in the terminology of Ref.~\cite{Altland1997}, obey the
symmetry
$$
\sigma_3 {\cal D} \sigma_3 = -{\cal D}.$$
Thus if $\psi$ is an
eigenvector of ${\cal D}$ with the eigenvalue $E$, $\sigma_3 \psi$ is
an eigenvector with the eigenvalue $-E$. Thus, (after a suitable
diagonalization) the zero modes of ${\cal D}$ must obey the relation
$$
\sigma_3 \psi_{L,R} = \pm \psi_{L,R}.$$
Namely, they are
eigenstates of the $\sigma_3$ operator, with the ``left" zero modes
$\psi_L$ coming with the eigenvalue $+1$, and the ``right" zero modes
$\psi_R$ labelled by the eigenvalue $-1$. As the operator ${\cal D}$
is deformed, the number of zero modes changes, but the non-zero modes
always appear in pairs, where one of the members of a pair has to be
``left" and the other ``right".  Therefore, while the number of zero
modes is not an invariant, the difference between the number of left
and right zero modes is a topological invariant, determined (through
the index theorem) by the monopole charge of the background
gauge-field.

Contrast this with zero modes of ${\cal H}$, which obey the relation
\rfs{eq:rela}. Because of the complex conjugation on $\psi$, these
zero modes cannot be split into ``left" and ``right". Indeed, even if
we tried to impose $\sigma_1 \psi^* = - \psi$, a simple redefinition
of $\psi \rightarrow i \psi$ brings this relation back to
\rfs{eq:rela}. Thus, the most an ``index theorem" could demonstrate in
the case of the BdG problem, is whether there is 0 or
exactly 1 zero mode. Moreover, since the BdG problem
is defined by an even-dimensional Hamiltonian, generically there will
not be any topologically protected zero
modes~\cite{NickReadUnpublished,ZirnbauerPrivate}.

Yet it is quite remarkable that in the case of an isolated vortex of
odd vorticity in a macroscopic sample (i.e., ignoring the boundaries)
of a $p_x+ip_y$ superconductor of spinless fermions, there is exactly
one zero mode localized on this vortex
\cite{Kopnin1991,Read2000,Nayak2006a,Tewari2006}. To be consistent
with above general property of the BdG Hamiltonian
(namely, that the total number of zero modes must be even) another
vortex is situated at the boundary of the system
\cite{Read2000,NickReadUnpublished}, preserving the overall parity of the number of
zero modes. Hence, although even in this odd-vorticity case the
one zero mode is not protected topologically, able to hybridize with a
vortex at a boundary of the sample, it survives (up to exponentially
small corrections) only by virtue of being far away from the boundary (and from other
odd-vorticity vortices).

To see this explicitly we consider the BdG equations
for a two-dimensional $p_x+ip_y$ superconductor
 \begin{eqnarray} \label{eq:BdGplane}
\left(- \frac{\nabla^2}{2m} - \mu \right) u(\r)- \sqrt{\Delta(\r)} \pp{\bar z} \left[ v(\r) \sqrt{\Delta(\r)}
  \right] = E u(\r), \cr \left( \frac{\nabla^2}{2m} + \mu \right)
v(\r)- \sqrt{ \Delta^*(\r)}  \pp{z} \left[
u(\r)  \sqrt{\Delta^*(\r)}  \right] = E v(\r) .
\end{eqnarray}
Here $\Delta(\r)$ is the order parameter of the superconductor,
$z=x+iy$, $\bar z=x-iy$ are the two-dimensional complex coordinates,
$m$ is the fermion mass, and $\mu$ is the chemical potential.
\rfs{eq:BdGplane} is of course a particular case of the eigenvalue
equation for a matrix of the form given in \rfs{eq:classD}, with the
vector $\psi$ represented by
\begin{equation} \label{eq:defpsi}
  \psi = \left( \matrix{ u \cr v }\right).
\end{equation}
For a uniform (vortex-free) order parameter, $\Delta(\r)=\Delta_0$, it
is easy to solve \rfs{eq:BdGplane} in terms of plane waves, finding
the spectrum
\begin{equation} \label{eq:spectrum}
E_{\k } = \sqrt{ \left(\frac{k^2}{2m} - \mu \right)^2
+ |\Delta_0|^2 k^2}.
\end{equation}
Since $E_{\k}$ has a gap for all $\k$ (with the exception of the
critical point at $\mu=0$~\cite{Read2000,VolovikBook}), consistent with
above discussion, there are no zero modes of \rfs{eq:BdGplane} in the
absence of vortices.

Now consider a superconductor with a symmetric vortex of vorticity
$\ell$. The order parameter is then given by
\begin{equation}
\Delta(\r)= e^{i\ell \varphi} f^2( r),
\end{equation}
where $r$, $\varphi$ are the polar coordinates centered on the vortex
and $f(r)$ is a real function of $r$ that vanishes at small $r$. Then
the BdG equations take the form
\begin{eqnarray} \label{eq:BdGvortex}
\left(- \frac{\nabla^2}{2m} - \mu \right) u- f(r)
e^{\frac{i\ell\varphi}{2}} \pp{\bar z} \left[  e^{\frac{i\ell\varphi}{2}}
f(r) v \right] = E u, \cr \left( \frac{\nabla^2}{2m} + \mu \right)
v- f(r)  e^{- \frac{i\ell\varphi}{2}} \pp{z} \left[
e^{-\frac{i\ell\varphi}{2}}  f(r)  u \right] = E v.
\end{eqnarray}

Next we observe that for the case of a vortex of {\em even} vorticity,
$\ell=2n$, we can eliminate the phase dependence of \rfs{eq:BdGvortex}
entirely . Indeed, making a transformation
\begin{equation} \label{eq:cltr}
u \rightarrow u e^{i n\varphi}, \  v \rightarrow v e^{-i n \varphi}.
\end{equation}
leads to equations
\begin{eqnarray} \label{eq:BdGvortex1}
\left(- \frac{\nabla^2}{2m} +\frac{n^2}{2m r^2}- \mu \right) u-
\frac{ i n}{m r^2} \pbyp{u}{\varphi}- f(r)  \pp{\bar z} \left[  f(r)
v \right] = E u, \cr \left( \frac{\nabla^2}{2m} -\frac{n^2}{2m r^2}+
\mu \right) v- \frac{ i n}{m r^2} \pbyp{v}{\varphi}-f(r)  \pp{z}
\left[  f(r)  u \right] = E v. \cr
\end{eqnarray}
Now we note that these equations are topologically equivalent to the
BdG equations without any vortices. Indeed, the only difference between
these equations and those for a uniform condensate is the presence of
the terms $2 i n/r^2[\partial/\partial \varphi]$, $n^2/r^2$, and
$f(r)$ that is a constant at large $r$ and vanishes in the core of the
vortex for $r<r_{\rm core}$.  We can imagine smoothly deforming these
equations to get rid of the first two terms (for example, by replacing
them with $\alpha \left(n^2/r^2 - 2 i n/r^2[\partial/\partial \varphi]
\right) u$ and taking $\alpha$ from $1$ to $0$), and smoothly
deforming $f(r)$ into a constant equal to its asymptotic value at
large $r$; in order to be smooth, the deformation must preserve the
BdG structure \rfs{eq:classD} and the vorticity of the order
parameter.  These equations then become equivalent to
\rfs{eq:BdGplane} for a constant, vortex-free order parameter
$\Delta(\r)=\Delta_0$ with an exact spectrum \rfs{eq:spectrum}, that
for $\mu\neq 0$ clearly does not exhibit any zero modes.

As Eqs.~\rf{eq:BdGvortex1} are smoothly deformed, 
in principle it is
possible that for a particular 
deformation some of its
eigenstates will become zero modes (although, as demonstrated above,
this can only happen in $\pm E$ pairs, leading to an even number of
these).  However, these modes will not be topologically protected, and
even a small deformation of, say, the order parameter shape $f(r)$
will destroy these modes. We note that this argument easily
accommodates vortices that are not symmetric, as those can be smoothly
deformed into symmetric ones without changing the topologically
protected parity of $N_z$.  The conclusion is that generically there
are no zero modes in the presence of an isolated vortex of even
vorticity.

In fact, if any doubts remain, it is also possible to directly
demonstrate the absence of zero modes in \rfs{eq:BdGvortex1}, simply by
following the arguments parallel to those given after \rfs{eq:ansatz}.
However, the arguments presented above are more general and robust, and
can be used to establish the claim even for non-symmetric
even-vorticity vortices.

The situation is drastically different if the vorticity of a vortex is
{\em odd}, i.e., $\ell=2n-1$. Indeed, in that case the transformation
\rfs{eq:cltr} cannot entirely eliminate such vortex from the equations
(even with the help of a smooth deformation), leaving at least one
fundamental unit of vorticity. This thereby guarantees at least one
zero mode localized on the odd-vorticity vortex.  To see this, recall
that due to the condition \rfs{eq:rela} together with the definition
\rfs{eq:defpsi}, the zero mode satisfies
\begin{equation}
u=v^*.
\end{equation}
Combining this with the transformation \rfs{eq:cltr}, we
find the equation for the zero mode
\begin{eqnarray} -  f(r) e^{-\frac{i\varphi}{2}} \pp{\bar z} \left[  e^{-\frac{i \varphi}{2}}   f(r) u^* \right]  &=& \cr
\left(\frac{\nabla^2}{2m} - \frac{n^2}{2m r^2}+ \mu \right) u+\frac{
i n}{m r^2} \pbyp{u}{\varphi}.  && \end{eqnarray} We look for a
solution to this equation in terms of a spherically symmetric real
function $u(r)$. This gives
\begin{equation} \label{eq:shrspsym}
-\frac{1}{2m} u'' - \left(  \frac{f^2}{2} + \frac{1}{2m r} \right)
u' - \left( \frac{f^2}{4r} + \frac{f f'}{2} -\frac{n^2}{2mr^2}
\right) u = \mu u.
\end{equation}
A transformation
\begin{equation} \label{eq:mcs}
u(r) = \chi(r) \exp \left( - \frac m 2 \int_0^r dr'~ f^2(r') \right)
\end{equation}
brings this equation to the more familiar form
\begin{equation}
-\frac {\chi'' } {2m} - \frac{\chi' }{2mr} + \left( m
\frac{f^4(r)}{8} + \frac{n^2}{2mr^2} \right) \chi = \mu \chi.
\end{equation}
This is a Schr\"odinger equation for a particle of mass $m$ which
moves with angular momentum $n$ in a potential $m f^4/8$ that is
everywhere positive. We observe that this potential vanishes at the
origin, and quickly reaches its asymptotic bulk value $f_0$ away from
the origin. Then for $\mu>mf_0^4/8$, there always exist a solution to
this equation finite at the origin and at infinity. Moreover, if
$\mu<mf_0^4/8$, then the solution finite at the origin will diverge at
infinity as
\begin{equation}
\chi \sim e^{r\sqrt{m^2 \frac{f_0^4}{4}-2m\mu}}.
\end{equation}
Combining this with \rfs{eq:mcs}, we observe that $u(r)$ remains a
bounded function at infinity as long as $\mu>0$. Thus the conclusion
is, there exist a zero mode as long as $\mu>0$. For the special case
of the $n=0$ vortex of vorticity $-1$, the small and large $r$
asymptotics of the solution we found here was discussed recently in
Ref.~\cite{Nayak2006a}.

In the simplest London approximation of a spatially uniform
condensate with $f(r)=f_0$ for all $r$ except inside an infinitesimally small
core, the zero mode localized
on an isolated odd-vorticity
vortex is simply given by
\begin{eqnarray}
\hspace{-0.51cm}
\label{f0zeromode}
u(r)&=&\left\{ \matrix{  J_n\left(r \sqrt{2\mu m -m^2 \frac{f_0^4}{4}}
\right) e^{-\frac{m}{2} f_0^2 r}, \ {\rm for}~\mu>m \frac{f_0^4}{8},
\cr I_n\left(r \sqrt{m^2 \frac{f_0^4}{4}-2m \mu} \right)
e^{-\frac{m}{2} f_0^2 r},  \ {\rm for}~0<\mu<m \frac{f_0^4}{8} , }
\right.
\nonumber\\
&&
\end{eqnarray}
where $J_n(x)$, $I_n(x)$ are Bessel and modified Bessel functions.

We note that it may seem possible to construct additional zero modes
in the following way. Instead of the ansatz of a rotationally
invariant $u(r)$ just before \rfs{eq:shrspsym}, we could have chosen
an ansatz
\begin{equation} \label{eq:ansatz}
u(r,\varphi)
= u_\alpha(r) ~e^{i\alpha \varphi} + u_{-\alpha}(r) ~ e^{-i \alpha \varphi}.
\end{equation}
Then two second order differential equations follow relating these
two functions.
Generally there are four solutions to these equations. Boundary
conditions at the origin $r=0$ select a subset of two of these
solutions.
Boundary conditions at infinity select a different subset of two solutions.
%
%
%
However, barring a coincidence, none of those solutions finite at the
origin are also finite at infinity.  Even if such coincidence arises
for some special value of $\mu$, by the above arguments, the
additional zero modes must appear in topologically unprotected pairs,
that will be split to finite $\pm E$ energies by a slight generic
deformation of the potential (order parameter distortion). Hence we
conclude that generically there will be no additional zero modes (except the one
found above) for an odd-vorticity vortex.


Thus we indeed find that the number of zero modes in a symmetric odd-vorticity vortex must be
one. Since a smooth deformations of the order parameter  can only change the zero mode number by multiples of two, 
an arbitrarily shaped odd-vorticity vortex must have an odd number of zero modes. However, any number of zero modes other than one is not generic and will  revert to one under an arbitrary deformation of the order parameter.  


To summarize, the results presented here establish the robustness of the zero modes
localized on well-separated ($r_{\rm separation} \gg 1/(m\Delta)$) odd-vorticity vortices and support the idea that they can
eventually be used to demonstrate non-Abelian statistics and perhaps
even for quantum computation.

This work was supported by the NSF DMR-0449521 (V.G.), DMR-0321848 (L.R.).
The authors are grateful to Martin Veillette for many useful
discussions and especially to Martin Zirnbauer for clarifying the
meaning of the index theorem for the Bogoliubov-de-Gennes
Hamiltonians.

\bibliography{zero}

\end{document}